\documentclass{elsarticle}
\makeatletter
\def\ps@pprintTitle{%
 \let\@oddhead\@empty
 \let\@evenhead\@empty
 \def\@oddfoot{}%
 \let\@evenfoot\@oddfoot}
\makeatother
\usepackage{geometry}                
\usepackage{graphicx}
\usepackage{color, xcolor}
\usepackage{amsmath,amssymb}
\usepackage{url}

\newcommand{\beq}{\begin{equation}}
\newcommand{\eeq}{\end{equation}}
\newcommand{\beqr}{\begin{eqnarray}}
\newcommand{\eeqr}{\end{eqnarray}}

\begin{document}

\title{Cohomology ring of the BRST operator associated to the sum of two pure spinors}

\author[andrei]{A. Mikhailov}
\author[UCDavis-Mat]{A. Schwarz}
\author[UCDavis-Phy]{Renjun Xu\corref{cor}}
\ead{rxu@ucdavis.edu} \cortext[cor]{Corresponding author.}
\address[andrei]{Instituto de F\'{i}sica Te\'orica, Universidade Estadual Paulista\\  R. Dr. Bento Teobaldo Ferraz 271, Bloco II -- Barra Funda\\  CEP:01140-070 -- S\~{a}o Paulo, Brasil}
\address[UCDavis-Mat]{Department of Mathematics, University of California, Davis, CA 95616, USA}
\address[UCDavis-Phy]{Department of Physics, University of California, Davis, CA 95616, USA}

\begin{abstract}
In the study of the Type II superstring, it is useful to consider
the BRST complex associated to the sum of two pure spinors.
The cohomology of this complex is an infinite-dimensional vector space.
It is also a finite-dimensional algebra over the algebra of functions of
a single pure spinor. In this paper we study the multiplicative structure.
\end{abstract}

\maketitle

The central object of the pure spinor formalism is the BRST operator $Q$, which
involves nonlinearly constrained ghosts. For example, the Type IIB superstring
uses two ghosts $\lambda_L$ and $\lambda_R$, which are spinors of $so(10)$ satisfying the
{\em pure spinor constraint}:
\begin{equation}\label{PureSpinorConstraint}
   (\lambda_L\Gamma^m\lambda_L) = (\lambda_R\Gamma^m\lambda_R) = 0
\end{equation}
Roughly speaking, the BRST structures are in one-to-one correspondence with
SUGRA backgrounds. For a given background, the cohomology of $Q$ describes
its infinitesimal deformations. It was shown in \cite{Berkovits:2012ps,Mikhailov:2013dp} that the knowledge of the
cohomology of the simplified differential $Q^{(0)}=(\lambda_L+\lambda_R)\frac{\partial}{\partial \theta}$ where $\theta$ is an
odd spinor variable is useful for understanding the cohomology of $Q$.

The cohomology groups of $Q^{(0)}$ were calculated in \cite{Mikhailov:2013dp} as modules over the ring
of polynomials of $\lambda_L^{\alpha},\lambda_R^{\alpha}$. In the present paper we will study the
multiplication of the cohomology classes, and calculate the cohomology as
a ring. This allows us to simplify the description of cohomology. We can use
the multiplication in cohomology to obtain information about  cohomology of
more complicated differentials.

Let us consider the ring $A=\mathbb{C}[\lambda_L,\lambda_R,\theta ]$ of polynomials depending of
ten-dimensional even pure spinors  $\lambda_L,\lambda_R$ and a ten-dimensional odd spinor $\theta.$
In other words, this means that  the components $\lambda_L,\lambda_R$ obey the pure spinor
constraint (\ref{PureSpinorConstraint}) and the components $\theta^{\alpha}$ are free Grassmann variables.

We define a differential acting on $A$ by the formula:
\begin{equation}
   Q =  \left(\lambda_L^{\alpha} + \lambda_R^{\alpha}\right)
\frac{\partial}{\partial \theta^{\alpha}}
\end{equation}
This differential commutes with the multiplication by a polynomial of $\lambda_L,\lambda_R$.
Let $R$ denote the ring of polynomials of $\lambda_L$ and $\lambda_R$:
\begin{equation}
   R =
{\bf C}[\lambda_L,\lambda_R]
\end{equation}
Then $A$ is a DGA over $R$, {\it i.e.} a differential graded $R$-algebra.
The grading is introduced as the number of thetas (we will use the term
$\theta$-degree for this grading). As $Q$ commutes with the elements of $R$, the
cohomology $H(A)$ can also be regarded as a graded $R$-algebra. The natural
action of $so(10)$ on $A$ induces a representation of $so(10)$ on $H(A).$ The
permutation of $\lambda _L$ and $\lambda_R$ commutes with the differential and therefore
induces an involution on $H(A).$  Our goal is to calculate $H(A)$  with all
of these structures.

First of all we describe the elements that generate $H(A)$ as a unital
$R$-algebra. This means that all cohomology classes can be obtained from them
and the unit element by means of multiplication, addition and multiplication
by a polynomial of $\lambda_L,\lambda_R.$

We prove that  $H(A)$ is generated by the cohomology classes of the following
elements:
\begin{align}
    \Phi^p & =
(\lambda_L - \lambda_R)\Gamma^p \theta \;, \quad\mbox{deg }\Phi^p=1,
\\[5pt]
\Phi \;& =
(\lambda_L\Gamma^p\theta)(\lambda_R\Gamma^p\theta)\;, \quad\mbox{deg } \Phi=2,
\\[5pt]
\Psi_{\alpha} & = \frac{\partial}{\partial\theta^{\alpha}}\Psi\;,
\quad\mbox{deg } \Psi_{\alpha}=4,
\\[5pt]
\Psi \;& =
\left(1 - \frac{5}{3}\left(\lambda_L 
\frac{\partial}{\partial\lambda_R}\right)\right)
(\lambda_R\Gamma^p\theta)(\lambda_R\Gamma^q\theta)(\lambda_R\Gamma^r\theta)
(\theta\Gamma_{pqr}\theta) \; - \;\{R\to L\}\;,
\label{DefPsi}\\
& \quad\mbox{deg } \Psi=5
\nonumber
\end{align}
Here   $\frac{\partial}{\partial\lambda_R}$ stands for a formal differentiation with respect to $\lambda_R$, without
taking into account the pure spinor constraint. The second term in (\ref{DefPsi}) is
obtained from the first one by interchanging $L$ and $R.$

Using the algebra  generators we can write down elements that generate $H(A)$ as an $R$-module.
Namely, we should add to them the unit element of the algebra,  $45$ elements $\Phi^{ab}$ and $120$ elements   $\Phi^{abc}$ given by formulas
\begin{equation}
\label{ }
\Phi^{ab}=\Phi^a\Phi^b
\end{equation}
\begin{equation}
\Phi^{abc}=\Phi^a\Phi^b\Phi^c
\end{equation}
In other words, all elements of $H(A)$ can be obtained as linear combinations of $1,\Phi^p,\Phi,$ $\Psi_{\alpha},$ $\Psi,$ $\Phi^{ab},$ $\Phi^{abc}$ with coefficients
belonging to $R=\mathbb{C}[\lambda_L,\lambda_R]. $ The $so(10)$-action on $H(A)$ is determined by the action on ring generators.  The generators $\Phi$ and $\Psi$ are scalars; ten generators $\Phi^a$ as a vector $V=[1,0,0,0,0]$;
sixteen generators $\Psi_{\alpha}$ transform according the dual spinor representation $S^*=[0,0,0,1,0]$.  It follows  that $\Phi^{ab}$ transforms as $\Lambda^2V=[0,1,0,0,0]$ and $\Phi^{abc}$ as $\Lambda^3V=[0,0,1,0,0]$

We can consider $H(A)$ as a vector space, or as a representation of $so(10)$. We can introduce an additional grading as the number of lambdas.
We denote:
\begin{itemize}
\item $H^{N,q}(A)$ the component of $H(A)$
   with $\lambda$-grading $(N-q)$ and $\theta$-grading $q.$ (Hence $N$ stands for total number of lambdas and thetas.)
\item $H^q(A)$ the component of $H(A)$  consisting of elements of $\theta$-degree $q$ ({\it i.e.} the sum of $H^{N,q}(A)$ over all possible $N$).
\end{itemize}
We can analyze the structure of $H^{N,q}(A)$ as $so(10)$-representation. We obtain
\beqr
H^{N,0}&=& [0,0,0,0,N] \label{Hk,0}\\
H^{N,1}&=& [1,0,0,0,N-2] \\
H^{N,2}&=& [0,1,0,0,N-4], \text{ when $N \neq 4$}\\
H^{N,3}&=& [0,0,1,0,N-6] \\
H^{N,4}&=& [0,0,0,1,N-7] \\
H^{N,5}&=& [0,0,0,0,N-8] \label{Hk,5}
\eeqr
When $N=4$, there is one additional term, a scalar, in $H^{4,2}$:
\beq
H^{4,2}=[0,0,0,0,0] \oplus [0,1,0,0,0] \label{H4,2}
\eeq

Notice that the involution  of $H(A)$ corresponding to the interchange of $\lambda_L$
and $\lambda_R$ acts as a multiplication by $(-1)^{N-q}$  on almost all elements of
$H^{N,q}$, the only exception being $[0,0,0,0,0]$ in $H^{4,2}$ where it acts as the
multiplication by $-1$.  Sometimes it is more convenient to use the involution
$\lambda_L\to -\lambda _R$, $\lambda_R\to -\lambda_L$ combining the interchange of lambdas and the sign
change; the groups $H^{N,q}$ with the exception of $[0,0,0,0,0]$ in $H^{4,2}$  are
invariant with respect to this involution.

The vector space spanned by the generators of the $R$-module to $H^q(A)$ will
be denoted by $W^q$; this space carries a representation of $so(10).$  This
representation is irreducible in all cases except for $q=2$.  It is easy to
check that:
$$W^1=H^{2,1}=[1,0,0,0,0],$$
$$W^2=H^{4,2}=[0,0,0,0,0] \oplus [0,1,0,0,0] =W^2_1+W^2_2,$$
$$W^3=H^{6,3}=[0,0,1,0,0],$$
$$W^4=H^{7,4}=[0,0,0,1,0],$$
$$W^5=H^{8,5}=[0,0,0,0,0].$$
Some elements of $R$ act as zero on $H(A)$. Namely, consider the ideal of $R$
consisting of the polynomials of the form $(\lambda_L^{\alpha} + \lambda_R^{\alpha})P_{\alpha}(\lambda_L,\lambda_R)$ where $P_{\alpha}(\lambda_L,\lambda_R)$
can be any polynomial. This ideal acts as zero on $H(A)$, because:
\begin{equation}
(\lambda_L^{\alpha} + \lambda_R^{\alpha})P_{\alpha}(\lambda_L,\lambda_R)
= Q\left(\theta^{\alpha}P_{\alpha}(\lambda_L,\lambda_R) \right)
\end{equation}
Let us denote $R_+$ the factorspace of $R$ over this ideal. The degree $k$ component
$R_+^k$ carries the representation $[0,0,0,0,k]$ of $so(10).$ Geometrically, we can
interpret $R_+$ as the algebra of functions of a single pure spinor. We see that
$H^q(A)$ is actually an $R_+$-module. In other words, we have a map
\begin{equation}
a\;:\;R_+^k\otimes W^q\to H^{k+q+l(q),q}
\end{equation}
where $l(q)$ stands for the $\lambda$-degree of $W^q.$ One can check that this map is
surjective  (moreover, for $q=2$ this statement remains correct if $W^{q=2}$ is
replaced with $W^2_2$; in fact $R^{k>0}_+\otimes W^2_1\subset \mbox{ker }a$). In other words, all
cohomology classes can be obtained as linear combinations of generators with
coefficients from $R_+.$

Using the above results we can find the bases of vector spaces $H^{N,q}(A).$ For example, one can find  the highest weight vectors of the representations acting on these spaces; then one can construct a basis applying the operators corresponding to  negative roots.
The highest weight vector  of $H^{N,q}$ can be obtained as a product of the highest weight vector of  the $k$-th graded component of the ring $R_+$ and the highest weight vector of $W^q$. (Here $k=N-q-l(q)$.)
For example, when $q=2$, the highest weight vector of $H^{N,2}$ can be obtained as a product of $(\lambda_L^1 - \lambda_R^1)^{N-4}$ and the vector $\Phi^1\Phi^2$ in $W^2.$ {\footnote{Here we use a convention of choosing $\lambda^1$ and $\Phi^1\Phi^2$ as the highest weight vectors  in spinor representation and in $\Lambda^2V$ respectively}}

Our results can be applied to the analysis of the cohomology of the differential
$Q_{ch}$ that is obtained from $Q$ adding the expression $((\lambda_L - \lambda_R)\Gamma^p \theta)\frac{\partial}{\partial x^p}.$ (This operator acts in the space of polynomial functions of $\lambda_L,\lambda_R, \theta $ and $x$. One can consider instead of polynomials of $x$ other classes of functions.) Using the notations above we can represent $Q_{ch}$ in the form

$$Q_{ch}=Q+\Phi^p\frac{\partial}{\partial x^p}.$$

To calculate the cohomology of $Q_{ch}$ we can apply the spectral sequence of bicomplex (in more physical words, we can consider the second summand as a perturbation). The first approximation
(the $E_2$-term of spectral sequence ) is given by the cohomology of the differential induced by the second summand on the cohomology of the first summand (on $H(A)\otimes \mathbb{C}[x]$).
To compute this differential we should calculate the product of $\Phi^p$ with generators of modules $H^q(A)$. We obtain
\begin{align}
\Phi^p \Phi^q =\;& \Phi^{pq}
\\
\Phi^p \Phi =\;& 0
\\
\Phi^p\Phi^{ql} =\;& \Phi^{pql}
\\
\Phi^p\Phi^{qlm} \simeq \;& (\Gamma^{pqlm})^{\alpha}_{\beta}
(\lambda_L^{\beta} - \lambda_R^{\beta})\Psi_{\alpha}
\label{Phi1Phi3}
\\
\Phi^p\Psi_{\alpha} \simeq \;&
\Gamma ^p_{\alpha \beta}(\lambda_L^{\beta}-\lambda_R^{\beta})\Psi
\label{Phi1Psi}
\\
\Phi^p\Psi =\;& 0
\end{align}
We have not calculated the coefficients in (\ref{Phi1Phi3}) and (\ref{Phi1Psi}).

We can calculate also the complete multiplication table of generators of
$H(A)$ considered as $R$-module.

Let us describe the way we obtained our results. Using the Macaulay2 package \cite{mac2} we can calculate the number of generators of the $R$-algebra $H(A)$,  the degrees of these generators, and the number of generators   of the $R$-modules $H^k(A)$.  We find that  the $R$-algebra $H(A)$  is generated by $28$ elements (ten of degree $1$, one of degree $2$, sixteen of degree $4$, and one of degree $5$), the number of generators of $H^q(A)$ is equal to $195$ (with one of degree $0$, ten of degree $1$, $46$ of degree $2$, $120$ of degree $3$, sixteen of degree $4$, and one of degree $5$). The calculation gives us also the Hilbert series of $H(A)$ (the generating function of $dim H^{N,q}$); we obtain
$$\sum dim H^{N,q}\tau_0^{N}\tau_1^q=\displaystyle\sum_{q=0}^5 P_q(\tau_0)\tau_1^q,$$
where
\begin{eqnarray*}
P_0(\tau_0)&=&\frac{1 + 5 {\tau_0} + 5 {\tau_0}^2 + {\tau_0}^3}{(1 - {\tau_0})^{11}}\\
P_1(\tau_0)&=&\frac{10 {\tau_0}^2 + 34 {\tau_0}^3 + 16 {\tau_0}^4}{(1 - {\tau_0})^{11}}\\
P_2(\tau_0)&=&(46 {\tau_0}^4 + 54 {\tau_0}^5 + 66 {\tau_0}^6 - 166 {\tau_0}^7 + 330 {\tau_0}^8 - 462 {\tau_0}^9 + 462 {\tau_0}^{10} - 330 {\tau_0}^{11} \nonumber\\
 && + 165 {\tau_0}^{12} - 55 {\tau_0}^{13} + 11 {\tau_0}^{14} - {\tau_0}^{15})/{(1 - {\tau_0})^{11}}\\
P_3(\tau_0)&=&(120 {\tau_0}^6 - 120 {\tau_0}^7 + 330 {\tau_0}^8 - 462 {\tau_0}^9 + 462 {\tau_0}^{10} - 330 {\tau_0}^{11} + 165 {\tau_0}^{12} \nonumber\\
 && - 55 {\tau_0}^{13} + 11 {\tau_0}^{14} - {\tau_0}^{15})/{(1 - {\tau_0})^{11}}\\
P_4(\tau_0)&=&\frac{16 {\tau_0}^7 + 34 {\tau_0}^8 + 10 {\tau_0}^9}{(1 - {\tau_0})^{11}}\\
P_5(\tau_0)&=&\frac{{\tau_0}^8 + 5 {\tau_0}^9 + 5 {\tau_0}^{10} + {\tau_0}^{11}}{(1 - {\tau_0})^{11}}\\
\end{eqnarray*}
To find the decomposition of $H^{N,q}$ into direct sum of irreducible representations of $so(10)$  we use the Lie code \cite{LiEcode} that allows to find such a decomposition for the graded components $A^{N,q}$ of $A$ for small $N,q.$ Applying Schur's lemma and the information about the Hilbert series we can find the action of the differential  on the irreducible components of $A^{N,q}$ ; this allows us  to find the $so(10)$-action on $H^{N,q}$ for small $N,q.$ (We use a version of the "maximal propagation principle"\cite{Movshev:2011pr},\cite{Cederwall:2001dx} in this consideration.) These results allow us to guess the representatives of the cohomology classes of generators. We can check our guess using \cite{mac2}.

Knowing the  $so(10)$-action on $H^{N,q}$
for small $N,q$ we guess the $so(10)$ representation for all $N,q$; we check our guess using the formula for the dimension of representation and the Hilbert series. To give a rigorous proof we can verify that the representation we guessed is embedded in $H^{N,q}$.

{\bf Acknowledgements}  We are indebted to N. Berkovits and M. Movshev for interesting discussions.


\def\cprime{$'$} \def\cprime{$'$}
\providecommand{\href}[2]{#2}\begingroup\raggedright\endgroup

\end{document}